\documentclass[11pt]{article}
\usepackage{graphicx,  epsfig}

\usepackage{amsmath,amssymb}

\newtheorem{defi}{Definition}[section]
\newtheorem{cor}{Corollary}[section]
\newtheorem{theo}{Theorem}[section]

\newtheorem{lem}{Lemma}[section]
\newtheorem{rema}{Remark}[section]

\newenvironment{dem}{\noindent \bf Proof: \rm}{$\quad\Box$}

\def \UN{\hbox{1\hskip -3.2pt I}}

\begin{document}
\title{Estimation of safety areas for epidemic spread}

\author{Beatriz Marr\'on $\, ^{*,1}$,  Ana Tablar  \begin{footnote}
{Departamento de Matem\'atica. Universidad Nacional del Sur. $^1$ Corresponding author: beatriz.marron@uns.edu.ar}
\end{footnote}}\date{May 10, 2001}
\maketitle

{\bf Abstract}\\In this work we study safety areas in epidemic spred. The aim of this work is, given the evolution of epidemic at time $t$, find a safety set at time $t+h$. This is, a random  set $K_{t+h}$ such that the probability that infection reaches $K_{t+h}$ at time $t+h$ is small.

More precisely, inspired on the study of epidemic spread, we consider a model in which the measure $\mu_n(A)$ is the incidence -density of infectives individuals- in the set $A$, at time $n$ and
$$\mu_{n+1}(A)(\omega)=\int_S{\pi_{n+1}(A;s)(\omega)\mu_n(ds)(\omega)}, \mbox {for any Borel set} \; A, $$
with random transition kernels of the form
$$\pi_n(.;.)(\omega)=\Pi(.;.)(\xi_n(\omega),Y_n(\omega)),$$
where $\xi$, $Y$ satisfy some ergodic conditions. The support of $\mu_n$ is called $S_n$. We also assume that $S_0$ is compact with regular border and that for any $x,y$
the kernel $\Pi(.;.)(x,y)$ has compact support.
A random set $K_{n+1}$ is a safety area of level $\alpha$ if:
\begin {itemize}
\item [{$i$)}] $K_{n+1}$ {\rm is a function of} $S_0, S_1, \cdots,S_n.$

\item [{$ii$)}] $P(K_{n+1} \cap S_{n+1} \neq \emptyset)\leq \alpha.$
\end{itemize}
We present a method to find these safety areas and some related results.\\


\section{Introduction}
Mathematical modelling for some type of epidemic spread, like those that affect the whole planet called \lq\lq pandemic", \, shoud reproduce two basic aspects observed in real data:
\begin{itemize}
\item [{a)}] A susceptible individual may be infected by individual who is usually located at a very distant point. For instance: an infected tourist infects someone he visits.
\par
\item [{b)}] The temporal evolution is typically non-markovian. Given the present, if the past indicates that the spread is diffusive or if the past shows that spread is in contraction, we will probably not make the same prediction for the future. That means that given the present, past and future may not be independent, which is a clear argument against Markov models.
\end{itemize}

In order to describe more precisely our model we introduce some basic notation: we denote by ${\cal S}$ a space of sites (typically ${\cal S}=\mathbb{Z}^d$, $\mathbb{R}^d$ or a finite subset of $\mathbb{R}^d$), by ${\cal B}({\cal S})$ its Borel $\sigma$-algebra and by ${\cal P}({\cal S})$ the set of probability measure on $({\cal S},{\cal B}({\cal S}))$.

A transition probability kernel (TPK) on ${\cal S}$, $\pi (A;s)$, is a function
 $$\pi:{\cal B}({\cal S}) \times {\cal S} \longrightarrow [0,1],$$
such that:
\begin{itemize}
\item For any fixed  $ s \in {\cal S}$, $\pi (.; s) \in {\cal P}({\cal S}).$
\item For any fixed $A \in {\cal B}({\cal S})$, $\pi (A;s)$ is a ${\cal B}({\cal S})$-measurable function.
\end{itemize}

The set of TPK on $S$ is denoted by ${\cal K}({\cal S})$, and in this work we deal with random TPK that describe probabilities for infection from one point to another. Loosely speaking, $\pi (A;s)$ gives the probability of a transition from $s$ to an element of $A$, in this contex, \lq\lq transition\rq\rq $\;$  means \lq\lq infection\rq\rq.
The state of the epidemic propagation at time $t = 0$ is described by a non-random probability measure $\mu_0$, so $\mu_0(A)$ gives the infection density on $A$ for any Borel set $A$, this is the number of infected people living in $A$  /total number of infected people. If $\mu_n$ describes the state at time $t = n$, then the description at time $t = n+1$ is given by

\begin{equation}
\mu_{n+1}(A)(\omega )= \int_S \pi_{n+1}{\rm (A;s)}(\omega )\mu_n (ds)(\omega ),
\label{mu}
\end{equation}
where $\pi_1(.;.)(\omega),\,\pi_2(.;.)(\omega),\cdots$ is a sequence of random TPK of the form
\begin{equation}
\pi_n(.;.)(\omega)=\Pi(.;.)(\xi_n(\omega),Y_n(\omega)), \; \mbox{for  any}\: n \: \mbox {and} \: \omega,
\label{pi}
\end{equation}
where
\begin{itemize}
\item [{$i$)}] $\Pi(.;.)(e,e^\prime):E \times E^\prime \longrightarrow {\cal K}({\cal S})$ is a measurable function, where $E$ and $E^\prime$ are polish spaces. We will assume in addition that this function is continuous on the second coordinate $e^\prime$.
\item [{$ii$)}] $\xi=(\xi_n)_{n\in {\rm \mathbb{N}}}$ is an {\it i.i.d} \, sequence of $E$-valued random variables.
\item [{$iii$)}] $Y=(Y_n)_{n\in N}$ is an $E^\prime$-valued process satisfying that its empirical measure $F^n_Y$ defined by $F^n_Y(B)(\omega)=\frac{\rm 1}{\rm n} \sum^n_{i=0} \UN_{\{Y_i(\omega)\in B \}}$ for any Borel set $B$ of $ E^\prime$, converge to a random measure $\lambda^Y(\omega)$ in this way: $$\sqrt n \sup_{x\in \mathbb{R}^+} \vert F^n_Y(x)(\omega)-\lambda^Y(x)(\omega)\vert\longrightarrow 0 \;\; {\mbox {a.s.}}$$
\item [{$iv$)}] $\xi$ and $Y$ are independent.
\end{itemize}

\begin{rema}
{\rm Let us note that, with the exception of $\mu_0$, $\mu_n$ are random measures.}
\end{rema}

\begin{rema}
{\rm For a random process $Y=(Y_n)_{n\in N}$ denote $\sigma^Y_m$ the $\sigma$-algebra generated by $\{Y_n :n \ge m\}$ and define
$$\sigma^Y_{\infty}=\bigcap^{\infty}_{m=1}{\sigma^Y_m},$$
we will say that $Y$ is regular if $\sigma^Y_{\infty}$ is trivial, in the sense that for any $A\in \sigma^Y_{\infty}$ one has $P(A)=0$ or $P(A)=1$. It is easy to check that the limit measure $\lambda^Y$ in ($iii$) is a  $\sigma^Y_{\infty}$-measurable random measure. Therefore, if $Y$ is regular, $\lambda^Y$ is deterministic, non-random.}
\end{rema}

The intuitive idea behind this type of models is that the evolution models of the type (\ref {mu}) allow to easily model transitions and to consider random transition kernels permits to model a complex and highly variable transition dynamics. For instance, in a pandemic spread, migration and touristic mobility play a key role on the spread. This currents of transition may have different \lq\lq regimens\rq\rq\,  with defferent intensities and directions.
Moreover, random TPK makes measures defined in (\ref{pi}) do not describe just a non-homogeneous Markov (as in the case when TPK are deterministic) but a non-Markov model, with dependence between past and future when present is given. Finally, the idea of descomposing randomness in two independient sources, one of them ($\xi$, corresponding to \lq\lq pure noise\rq\rq, in regresion analysis terms) of a  very simple probabilistic nature and the other ($Y$, corresponding to some \lq\lq explicative\rq\rq\,  variables, in regresion terms) possibly more complex but whose empirical measure converges, is a way to construct very general models where limits theorems can be easily obtained.

The aim of this paper is to find safety areas for the model (\ref{mu}), this is a random set such that the probability of an individual placed in that set to be infected in the next step is small.

We denote by $S_0$ the support of $\mu_0$, by $S_n$ the support of $\mu_n$, and by $d_i=diameter (S_i)$, and we will assume that
\begin{equation}
\label{bolas}
Support \, \left(\pi(\xi_i, Y_i) (.,s)\right)= B(s, r_i),\:
\mbox {for  any} \:s \in  S,
\end{equation}
where $B(s, r_i)$ is an open ball centered in s of radius $r_i$.

First, we find a safety area for the case that the radius $r_i$ depends only on $\xi_i$, while the whole measure $\pi(.,s)$ depends on both $\xi$ and $Y$, then we extend the definition for the case that the radius depends on $\xi$ and $Y$.

Some final remarks on general notation used all along this paper:
\begin{itemize}
\item $Z_n {\stackrel{w}{\longrightarrow}} Z$ or $Z_n {\stackrel{w}{\longrightarrow}} F$ denote a sequence of random variables $(Z_n)_{n\in N}$ that converges in distribution to a random variable $Z$ with distribution function $F$.
\item To simplify statements of results and definitions, we do not make explicit mention to obvious hypotheses: for instance, if a result refers to an integral, the integral is assumed to exist and be finite.
\item We indicate by \lq \lq :=\rq\rq\, a definition that is stated in the  middle of a formula.
\item Let $A^\delta$ denote the set $A^\delta=\{\,s\in S: distance(s,A)\le\delta\},$ for every $A \subset {\cal S}$.
\item Let $F_X^n(t)$ denote the empirical distribution function $F_X^n(t)=\frac{1}{n}\sum\limits_{i=1}^{n} \UN_{\{X_i\leq t\}},$ where $X_0, X_1, \cdots,X_n$ are random variables.
\end{itemize}

\section{Main results}
Let us consider here the case of ${\cal S}=\mathbb{R}^d$ or ${\cal S}=\mathbb{Z}^d$ (or, more in general, a subset of $\mathbb{R}^d$) and $E^\prime=\mathbb{R} $ (or, more in general, a subset of $\mathbb{R}$), $\mu_0$ a deterministic element of ${\cal P}({\cal S})$ and the sequence of random probability measures defined by (\ref{mu}). We assume that $S_0$ is compact and its border is a regular closed curve, an since $\mu_0$ is deterministic, so is $S_0$.

As we observe $\mu_0, \mu_1, \cdots,\mu_n$, then $d_0, d_1, \cdots,d_n$ are data, where $d_i=diameter (S_i)$, and by (\ref{bolas}), it is clear that $ d_{i+1}=d_i+2 \,r_{i+1}.$ Hence
\begin{equation}
r_{i+1}=\frac{d_{i+1}-d_i}{2}.
\label{r}
\end{equation}

This simple equation is basic for our purposes, because it means that we can compute $r_{i+1}$ in terms of $d_i$ and $d_{i+1}$, so we can compute
$r_1, r_2, \cdots,r_n$ in terms of $\mu_0, \mu_1, \cdots,\mu_n$, and we can base our statistical procedure on $r_1, r_2, \cdots,r_n$.

The intuitive concept of safety area given in the introduction, can be written more formally as follows, a safety area of level $\alpha$ is a random set $K_{n+1}$ that satisfies:
\begin{itemize}
\item [{$i$)}]  $K_{n+1}$ {\rm is a function of} $S_0, S_1, \cdots,S_n,$
\item [{$ii$)}]  $P(K_{n+1} \cap S_{n+1} \neq \emptyset)\leq \alpha$,
\end{itemize}
and the condition $ii)$ is equivalent to $P(K_{n+1} \cap S_n^{r_{n+1}}\neq \emptyset)\leq \alpha$, since $S_{n+1}=S_n^{r_{n+1}}$.

A simple way to find a safety area, is to choose  $\delta_{n+1}$, based on a sample of $r_1, r_2, \cdots,r_n$,  sufficiently large so that the set $K_{n+1}=(S_n^{\delta_{n+1}})^c$ satisfies the conditions $i)$ and $ii)$.

Let us note that   $\left \{(S_n^{\delta_{n+1}})^c \bigcap S_n^{r_{n+1}} \neq \emptyset \right\}$ and $\left\{r_{n+1} > \delta_{n+1} \right\}$ are equivalent events,  this is showed in the grafic in Figure 1.
\begin{figure}
\begin{center}
\epsfig{file=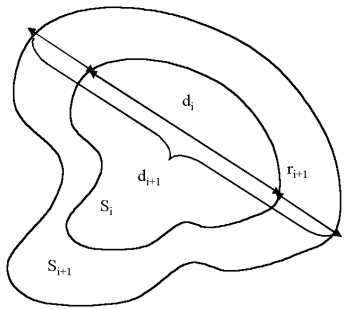,height=4cm}\hspace{1cm}
\caption{Safety area. }
\end{center}
\end{figure}

First we considere the case where each $r_i$ depends only on $\xi_i$ and, as $\xi_i$ are i.i.d., then $r_i=r(\xi_0).$ We call $F_0(x)$ the distribution function of the radius and we assume that $F_0$ is continuous. In this case, we define safety area, as follows.

\begin{defi}\label{def1}
A random set $K_{n+1}=(S_n^{\delta_{n+1}})^c$ is a safety area of level $\alpha$ if:
\item [{$i)$}]  $\delta_{n+1}$ {\rm is a function of} $r_1, r_2, \cdots,r_n,$
\item [{$ii)$}]  $P\left(r_{n+1} > \delta_{n+1} \right)\leq \alpha.$
\end{defi}

We present the following theorem, which provides a safety area under the conditions above.

\begin{theo}
The random set $K_{n+1}=(S_n^{\delta_{n+1}})^c$, where
$$\delta_{n+1}=\min \limits_{ t > 0} \left\{1-F_r^n(t)<\alpha - \mathbb{C}_n \right\}\;\; and \;\;\mathbb{C}_n=E \left(\sup \limits_{t \in\mathbb{R}^+}\left\vert F_r^n(t)-F_0(t) \right\vert\right),$$
defines a safety area of level $\alpha$.
\end{theo}

\begin{dem}

We need to prove
$P\left(r_{n+1} > \delta_{n+1}\right)\leq \alpha $.
Taking into account the fact that $r_{n+1}$ and $\delta_{n+1}$ are independent, we have

\begin{eqnarray*}
P\left(r_{n+1} > \delta_{n+1}\right)
&=& E\left(P\left(r_{n+1} > \delta_{n+1}/\,\delta_{n+1}\right)\right)\cr
&=& \int \limits_{0}^{\infty} P\left(r_{n+1}>u\right)\:dP^{\delta_{n+1}}\left(u\right)\cr
&=& \int \limits_{0}^{\infty} P\left(r_0>u\right)\:dP^{\delta_{n+1}}\left(u\right)\cr
&=& 1-\int \limits_{0}^{\infty}F_0\left(u\right)\:dP^{\delta_{n+1}}\left(u\right)\cr
&=& 1-E\left(F_0\left(\delta_{n+1}\right)\right),
\end{eqnarray*}
where $P^{\delta_{n+1}}$ is the distribution function of $\delta_{n+1}$.

We can write the last expresion as
 $$1-E(F_0(\delta_{n+1}))=1-E\left(F_r^n(\delta_{n+1})\right)+E \left(F_r^n(\delta_{n+1})- F_0(\delta_{n+1})\right),$$
 and since $0\le 1-F_r^n(\delta_{n+1})<\alpha-\mathbb{C}_n$ by definition, we only need to show that $\vert E \left(F_r^n(\delta_{n+1})- F_0(\delta_{n+1})\right) \vert \leq \mathbb{C}_n$.

Then,
\begin{eqnarray*}
\left\vert E \left(F_r^n(\delta_{n+1})- F_0(\delta_{n+1})\right) \right\vert
&\leq&E\left\vert F_r^n(\delta_{n+1})- F_0(\delta_{n+1})\right\vert\cr
&\leq&E\left(\sup\limits_{ t\in\mathbb{R}^+}\left\vert F_r^n\left(t\right)-F_0\left(t\right)\right\vert\right)\cr
&=&\mathbb{C}_n,
\end{eqnarray*}
and this completes the proof.
\end{dem}

Next we consider that the radius $r_i$ depends on $\xi_i$ and $Y_i$, so $r_i=r(\xi_i, Y_i)$.
Let $F(.;Y)$ be the  distribution of $r(\xi_0,Y)$, this is $F(t;Y)=P(r(\xi_0,Y)\leq t)$ for  any $t \ge 0,$ and let us suppose that $F(.;Y)$ is continuous.

We also assume that there exist a random probability distribution $F(x;\omega)$ such that
\begin {equation}
\sqrt n \sup_{t\in \mathbb{R}^+} \left\vert \frac{1}{n}\sum\limits_{i=1}^{n} F(t;Y_i(\omega))-F(t;\omega)\right\vert\mathop \rightarrow \limits_{n\to \infty}  0 \;\; {\mbox {a.s.}}
\label{F}
\end {equation}

In this case, we can not calculate the safety area in a straight way, then we define it using the limit distribution in (\ref{F}), as follows.

\begin {defi}
\label{def2}
A random set $K_{n+1}=(S_n^{\delta_{n+1}})^c$ is a safety area of level $(\epsilon, \alpha)$  if:
\begin {itemize}
\item [{$i$)}]  $\delta_{n+1}$ {\rm is a function of} $r_1, r_2, \cdots,r_n$,
\item [{$ii$)}] $\limsup_n P\left(1-F\left(\delta_{n+1};\omega\right) > \epsilon\right) \leq \alpha.$
\end {itemize}
\end {defi}

\begin{rema}
If $U_1,\cdots,U_n$ are independent random variables, uniformly distributed on $[0,1]$, it is well known that $\sup \limits_{t \in\mathbb{R}^+}\left\vert F_{U}^{n}- t \right\vert$,  has the law of the Kolmogorov-Smirnov statistic $\sup \limits_{t \in\mathbb{R}^+}\left\vert F_{X}^{n}-F(t)\right\vert$ for any $X_1,\cdots,X_n$ independent with a common distribution function $F(t)$. If $F(t)$ is continuous, by  Donsker Invariance Principle,
$$\sqrt n \; E \left(\sup \limits_{t \in\mathbb{R}^+}\left\vert F_X^n(t)-F(t) \right\vert\right) \mathop \rightarrow \limits_{n\to \infty} E\left(\sup\limits_{t\in[0,1]}\left\vert b\left(F(t)\right)\right\vert\right),$$
donde $b(t)$ es el puente Browniano.
\end{rema}

To prove the main result of this report, we will use the following theorem which proof is in the Appendix.
\begin {theo}
Suppose the random variables $X_1,\cdots,X_n$ are independent with continuous distribution function $F_1, \cdots, F_n$ and such that
\begin {itemize}
\item [{$i$)}] $\frac{1}{n}\sum\limits_{i=1}^{n}F_i(t)\mathop \rightarrow \limits_{n\to \infty}F(t).$
\item [{$ii$)}] $\frac{1}{n}\sum\limits_{i=1}^{n}F_i(s)\left(1-F_i(t)\right)\mathop \rightarrow \limits_{n\to \infty}G(s,t)$, positive and symmetrical function.
\item [{$iii$)}]$\limsup \frac{1}{n}\sum\limits_{i=1}^{n} w_i(\delta) \mathop \rightarrow \limits_{\delta \to 0^+}0,\; \mbox{where}\; w_i(\delta)\; \mbox{is the modulus of continuity of}\; F_i(t).$
\end{itemize}
Then the random variables $U_n$ defined by
$$U_n(t)=\sqrt {n} \left (F_X^n(t)-\frac{1}{n}\sum\limits_{i=1}^{n}F_i(t)\right),$$
satisfy $ U_n {\stackrel{w}{\Longrightarrow}} U$, where $U$ is the centered Gaussian process with covarience given by $E\left\{U(s)\,U(t)\right\}=G(s,t).$
\end{theo}

We assume in adition, that for a path fixed $y=(y_1,\cdots,y_n,\cdots)\in \mathbb{R}^{\rm \mathbb{N}}$,
$$\int \limits_{\mathbb{R}^{\rm \mathbb{N}}}C(\delta,y) dP^Y(y)\mathop \rightarrow 0 \;{\mbox {as}}\;\;
{\delta \to 0^+},$$
where $C(\delta,y)=\limsup_n\frac{1}{n}\sum\limits_{i=1}^{n}w(F(.;y_i),\delta).$

Under the conditions enumerated above, the next theorem provides a safety area for this case.

\begin{theo}
The random set $K_{n+1}=(S_n^{\delta_{n+1}})^c$ is a safety area of level $(\alpha, \epsilon)$,
taking $\; \delta_{n+1}=\min \limits_{ t > 0} \left\{\frac{1}{n} \sum\limits_{i=1}^{n} \UN_{\{r_i> t\}}<\epsilon-\displaystyle\frac{\mathbb{C}_{\alpha}}{\sqrt n}\right\} $, and $\mathbb{C}_{\alpha}$ verifying that $ P\left(\sup \limits_{t\in[0,1]}\left \vert U(t) \right \vert \geq \mathbb{C}_\alpha\right)=\alpha$.
\end{theo}

\begin{dem}
To prove that $K_{n+1}$ is a safety area, we need to show $(ii)$ in Definition \ref{def2}.

As
$$P\left(\left\vert 1-F(\delta_{n+1},\,.)\right\vert > \epsilon\right)\leq P\left(\left \vert \frac{1}{n} \sum \limits_{i=1}^{n}\left (\UN_{\{r_i \leq\delta_{n+1} \}}-F(\delta_{n+1},\,.)\right)\right \vert>\frac{\mathbb{C}_{\alpha}}{\sqrt n}\right)$$
$$+ P\left(\left\vert 1-\frac{1}{n} \sum \limits_{i=1}^{n} \UN_{\{r_i \leq \delta_{n+1}\}}\right\vert>\epsilon-\frac{\mathbb{C}_{\alpha}}{\sqrt n}\right),$$
from the definition of $\delta_{n+1}$ holds that the second term is equal to 0, then it is enough to prove that
\begin {equation}
\lim \limits_{n}P\left(\sup \limits_{t \in\mathbb{R}^+} \left\vert \frac{1}{n} \sum \limits_{i=1}^{n} \left(\UN_{\{r_i \leq t\}}-F(t;\omega)\right)\right \vert \geq \frac{\mathbb{C}_\alpha}{\sqrt n}\right)\leq\alpha.
\label{4}
\end{equation}

By (\ref{F}) and applying Lemma(\ref{lema1}) in Appendix, there exits a sequence of positive numbers $\{a_n\}$ such that $a_n\downarrow 0^+$ and
\begin {equation}
P\left(\sup_{t\in \mathbb{R}^+} \left\vert \frac{1}{n}\sum\limits_{i=1}^{n} F(t;y_i)-F(t;\omega)\right\vert>\frac{a_n}{\sqrt n}\right)\mathop \rightarrow \limits_{n\to \infty}  0.
\label{8}
\end{equation}
The argument of the limits in (\ref{4}) can be bounded in this way
\begin {eqnarray}
\label{an}
& &P\left(\sup \limits_{t \in\mathbb{R}^+} \left\vert \frac{1}{n} \sum \limits_{i=1}^{n} \left(\UN_{\{r_i \leq t\}}-F(t;\omega)\right)\right \vert \geq \frac{\mathbb{C}_{\alpha}}{\sqrt n}\right)\leq \;\;\;\;\;\;\;\; \;\;\;\; \;\;\;\;  \;\;\;\;   \;\; \;\;\;\;\;\;\;\; \;\;\;\; \;\;\;\;  \;\;\;\; \;\;\;\; \;\;\;\;   \;\;\;\;  \;\;\   \nonumber\\
& &P\left(\sup \limits_{t \in\mathbb{R}^+} \left\vert \frac{1}{n} \sum \limits_{i=1}^{n}\leq \left(\UN_{\{r_i \leq t\}}-F(t;y_i)\right)\right \vert \geq \frac{\mathbb{C}_{\alpha}-a_n}{\sqrt n}\right)+       \nonumber\\
& &   \;\;\;\;   \;\;\;\;  \;\;\;\;   \;\;\;\; \;\;\;\;\;\;\;\; \;\;\;\; \;\;\;\;  \;\;\;\;   P\left(\sup_{t\in \mathbb{R}^+} \left\vert \frac{1}{n}\sum\limits_{i=1}^{n} F(t;y_i)-F(t;\omega)\right\vert>\frac{a_n}{\sqrt n}\right),
\end {eqnarray}
and the second term in the last equation tends to 0, then we need to show that the first term tends to $\alpha$.

Let us fix a path $y=(y_1,\cdots,y_n,\cdots)\in \mathbb{R}^{\rm \mathbb{N}}$, and denote $\;{r^\prime}_i= r(\xi_i,y_i)$ with $i=1, \cdots,n$. Hence the radius ${r^\prime}_i$ are independent because they depend only on the $\xi_i$, then by an Invariance Principle applied to the Theorem \ref{empirica} $$\sqrt n \sup \limits_{t \in\mathbb{R}^+} \left \vert \frac{1}{n} \sum \limits_{i=1}^{n} \left (\UN_{\{r^{\prime}_i \leq t\}}-F(t;y_i)\right ) \right \vert \stackrel{w}{\Longrightarrow} \sup \limits_{t\in[0,1]}\left\vert U(t)\right\vert,$$
where $U(t)$ is a Gaussian centered process with autocovariance function $G(s,t)$.

By the assumption about $\mathbb{C}_\alpha$, we have
$$\lim \limits_{n}P\left (\sup \limits_{t \in\mathbb{R}^+} \left \vert \frac{1}{n} \sum \limits_{i=1}^{n} \left (\UN_{\{r^{\prime}_i \leq t\}}-F(t;y_i)\right ) \right \vert \geq \frac{\mathbb{C}_{\alpha}}{\sqrt n}\right )=\alpha.$$
\
This expression can be written as

$$\lim \limits_{n}P\left(\sup \limits_{t \in\mathbb{R}^+} \left\vert \frac{1}{n} \sum \limits_{i=1}^{n} \left(\UN_{\{r_i \leq t\}}-F(t;y_i)\right)\right \vert \geq \frac{\mathbb{C}_{\alpha}}{\sqrt n}\,/\,Y=y\right)=\alpha,$$
for all $y \in \mathbb{R}^{\rm \mathbb{N}}$ because the distribution of $r^{\prime}_1,\cdots,r^{\prime}_n, \cdots$ is just the same of  $r_1,\cdots,r_n, \cdots$ conditioned to $Y=y$.
Then
$$P\left(\sup \limits_{t \in\mathbb{R}^+} \left\vert \frac{1}{n} \sum \limits_{i=1}^{n} \left(\UN_{\{r_i \leq t\}}-F(t;y_i)\right)\right \vert \geq \frac{\mathbb{C}_{\alpha}}{\sqrt n}\right)=$$
$$\int \limits_{\mathbb{R}^{\rm \mathbb{N}}} P\left(\sup \limits_{t \in\mathbb{R}^+} \left\vert \frac{1}{n} \sum \limits_{i=1}^{n} \left(\UN_{\{r_i \leq t\}}-F(t;y_i)\right)\right \vert \geq \frac{\mathbb{C}_{\alpha}}{\sqrt n}\,/\,Y=y\right)dP^Y(y).$$

By Dominated Convergence Theorem, as the integrand tends to $\alpha$ and it is bounded between 0 and 1, we have that
\begin {equation}
\lim \limits_{n}P\left(\sup \limits_{t \in\mathbb{R}^+} \left\vert \frac{1}{n} \sum \limits_{i=1}^{n} \left(\UN_{\{r_i \leq t\}}-F(t;y_i)\right) \right\vert \geq \frac{\mathbb{C}_{\alpha}}{\sqrt n}\right)=\alpha. \label{lim}
\end{equation}

As $\alpha$ is any real number between 0 and 1, we have just proved that the equality
\begin {equation}
\lim \limits_{n}P\left(\sup \limits_{t \in\mathbb{R}^+} \left\vert \frac{1}{n} \sum \limits_{i=1}^{n} \left(\UN_{\{r_i \leq t\}}-F(t;y_i)\right)\right \vert \geq \frac{v}{\sqrt n}\right)=P\left(\sup \limits_{t \in [0,1]} \left\vert U_t \right\vert \geq v\right), \label{lim2}
\end{equation}
is valid for any real $v$.

If $G(t)$ is the distribution of the supreme of a Gaussian process $U$, by Lemma (\ref{lema2}) in Appendix,

$$\sup\limits_{v \in \mathbb{R}} \left\vert P\left(\sqrt n\sup \limits_{t \in\mathbb{R}^+}\left \vert \frac{1}{n} \sum \limits_{i=1}^{n} \left(\UN_{\{r_i \leq t\}}-F(t;\omega)\right)\right \vert \geq v\right)-\left(1-G(v)\right)\right\vert\mathop \rightarrow \limits_{n\to \infty}  0.$$

Then by (\ref{lim2})
\begin {eqnarray*}
\lim \limits_{n}P\left(\sqrt n \sup \limits_{t \in\mathbb{R}^+} \left\vert \frac{1}{n} \sum \limits_{i=1}^{n} \left(\UN_{\{r_i \leq t\}}-F(t;\omega)\right)\right \vert \geq \mathbb{C}_\alpha -a_n\right)
&=&\lim \limits_{n}P\left(\sup\limits_{t\in[0,1]}\left\vert U(t)\right\vert\geq \mathbb{C}_\alpha-a_n\right)\cr
&=&P\left(\sup\limits_{t\in[0,1]}\left\vert U(t)\right\vert\geq \mathbb{C}_\alpha\right)\\
&=&\alpha.
\end {eqnarray*}

Applying (\ref{8}) and the last equalities in (8), (6) follows.
\end{dem}

\section{Appendix}
\begin {theo}
Suppose the random variables $X_1,\cdots,X_n$ are independent with continuous distribution function $F_1, \cdots, F_n$ and such that
\begin {itemize}
\item [{$i$)}] $\frac{1}{n}\sum\limits_{i=1}^{n}F_i(t)\mathop \rightarrow \limits_{n\to \infty}F(t).$
\item [{$ii$)}] $\frac{1}{n}\sum\limits_{i=1}^{n}F_i(s)\left(1-F_i(t)\right)\mathop \rightarrow \limits_{n\to \infty}G(s,t),$ positive and symmetrical function.
\item [{$iii$)}]$\limsup \frac{1}{n}\sum\limits_{i=1}^{n} w_i(\delta) \mathop \rightarrow \limits_{\delta \to 0^+}0,\; \mbox{where}\; w_i(\delta)\; \mbox{is the modulus of continuity of}\; F_i(t).$
\end{itemize}
Then the random variables $U_n$ defined by
$$U_n(t)=\sqrt {n} \left (F_X^n(t)-\frac{1}{n}\sum\limits_{i=1}^{n}F_i(t)\right),$$
satisfy $ U_n {\stackrel{w}{\Longrightarrow}} U$, where $U$ is the centered Gaussian process with covarience given by $E\left\{U(s)\,U(t)\right\}=G(s,t).$
\end{theo}

\begin {dem}
We will derive this result by using the theory of week convergence in the space of continouos functions $\mathcal{C}$. Although $U_n(t)$ is a function on $[0,1]$ produced at random, it is not an element of  $\mathcal{C}$, being obviously discontinuous. Here we shall circumvent the discontinuity problems by adopting a diferent definition of empirical distribution function.

Let us define $X_{(0)}=-\infty$, $X_{(n+1)}=+\infty$ and $X_{(1)}, \cdots, X_{(n)}$ are the values $X_1, \cdots, X_n$ ranged in increasing order, and let $G_n(t)$ be the distribution function corresponding to an uniform distribution of mass $(n+1)^{-1}$over the intervals $(X_{(i-1)},X_{(i)}]$, for $i=2, \cdots, n$ and for the intervals $(-\infty; X_{(1)}]$ y $(X_{(n)}; +\infty)$ we assign the exponential distribution ${\cal E}(-\frac{ln(n+1)}{X(1)})$ and ${\cal E}(\frac{ln(n+1)}{X(n)})$ respectively.

Then
\begin{equation}
\left \vert F_X^n(t)-G_n(t) \right \vert \leq \frac{1}{n}, \; \; t \in \mathbb{R}.
\label{primera}
\end{equation}
Now let $Z_n$ be the element of $\mathcal{C}$ with value at $t$
\begin {equation}
Z_n(t)=\sqrt n \left( G_n(t)- \frac {1}{n}\sum\limits_{i=1}^{n} F_i(t)\right),
\label {0}
\end{equation}
and by (\ref{primera}) we have
\begin{equation}
\sup \limits_t \left \vert U_n(t)-Z_n(t)\right \vert \leq \frac{1}{\sqrt n}.
\label {sup}
\end{equation}

We really analyze $U_n$, replacing $U_n$ by $Z_n$ only in order to stay in $\mathcal{C}$, so we will prove
\begin{equation}
Z_n{\stackrel{w}{\longrightarrow}}U.
\label{Z}
\end{equation}

We will show first that the finite-dimensional distributions of $Z_n$ converge to those of   $U$. Consider a single time point $t$, we write
$U_n(t)= \sum\limits_{i=1}^{n} \varphi_i(t)$, where $\varphi_i(t)=\frac{1}{\sqrt n}\left(\UN_{\{X_i\leq t\}}-F_i(t)\right).$

Since $\left\{\varphi_i \right \}$ are a sequence of  centered, independent random variables  with variance $\sigma_i^2=\frac{1}{n}F_i(t)\left(1-F_i(t)\right)$ and such that $\vert \varphi_i\vert \leq \frac{1}{\sqrt n}$, for all $i$, by Chebyshev's inequality follows that

\begin{eqnarray*}
\sum_{\begin{array}{c}
        k \\
        \vert \varphi_i\vert \geq \epsilon \,S_n
      \end{array}}k^2 P\left(\varphi_i=k \right)
&\leq& \frac{1}{n}P\left\{\vert\varphi_i \vert \geq \epsilon \, S_n \right\}\\
&\leq& \frac{1}{n}\frac{{\rm var}(\varphi_i)}{\epsilon^2 \,S_n^2}\\
\end{eqnarray*}
where $S_n^2=\frac{1}{n}\sum \limits_{i=1}^{n}F_i(t)\left(1-F_i(t)\right)$ that has a finite limit, $G(t,t)$. Hence

$$ \frac{1}{S_n^2}\sum_{i=1}^{n} \sum_{\begin{array}{c}
        k \\
        \vert \varphi_i\vert \geq \epsilon \,S_n
      \end{array}}k^2 P\left(\varphi_i=k \right)\leq \frac{1}{n\,\epsilon^2 \,S_n^2}\rightarrow 0 \;\;\;\; as \;n\to \infty,$$
and the Lindeberg condition is satisfied, then follows that

$$U_n(t){\stackrel{w}{\longrightarrow}}N\left(0,G(t,t)\right).$$

Now consider two time point $s$ and $t$, with $s<t$. We must prove
$$a \,U_n(s)+ b\, U_n(t){\stackrel{w}{\longrightarrow}}N \left(0,a^2G(s,s)+b^2G(t,t)+ 2 a b G(s,t)\right),$$
for all $a,\,b \in \mathbb{R}$.

But $a \, U_n(s)+ b\, U_n(t)=\sum\limits_{i=1}^{n}\left(a \varphi_i(s)+b\varphi_i(t)\right)$,  and we denote $S_n^2$ the variance of $a \, U_n(s)+ b\, U_n(t)$.

Using  the fact that $\left\{a \varphi_i (s) + b \varphi_i (t) \right \}$ is a sequence of  centered, independent random variables  and such that $\vert a \varphi_i (s) + b \varphi_i (t)\vert \leq \frac{\vert a \vert + \vert b \vert}{\sqrt n}$, for all $i$, we see that

$$ \frac{1}{S_n^2}\sum_{i=1}^{n} \sum_{\begin{array}{c}k\\ \vert \varphi_i\vert \geq \epsilon \,S_n\end{array}}k^2 P\left(\varphi_i=k \right)\leq \frac{\left(\vert a \vert + \vert b \vert \right)^2}{n\,\epsilon^2 \,S_n^2}\rightarrow 0 \;\;\;\; as \;n\to \infty,$$

and the Lindeberg condition is satisfied.

Since $E\left\{U_n(s),U_n(t)\right\}=\frac{1}{n}\sum\limits_{i=1}^{n} F_i(s)\left(1-F_i(t)\right)$, $\left( U_n(s),U_n(t)\right)$ converge in law to a centered, bivariate normal such that $E\left\{U(s),U(t)\right\}=G(s,t).$

A set of three or more points can be treated in the same way, and hence the finite dimensional distribution of $U_n$ and, by (\ref{sup}), the finite-dimensional distributions of the $Z_n$ converge properly.
If we prove that $\{Z_n\}$ is tight, (\ref{Z}) will follow.

To prove the tightness of $\{Z_n\}$  it is enough to show that for each positive $\epsilon$ and $\eta$ there exists $\delta$,  $0 \leq \delta \leq 1$ and $\eta \geq \eta_0$ such that
$$P\left\{\sup \limits_{t \leq s \leq {t+\delta}}\left\vert Z_n(s)-Z_n(t)\right\vert \geq \epsilon\right\} \leq \delta\eta,$$
by (\ref{sup}) we need to prove that

\begin {equation}
P\left\{\sup \limits_{t \leq s \leq {t+\delta}}\vert {U_n(s)-U_n(t)}\vert \geq \epsilon\right\} \leq \delta\eta.
\label {2}
\end {equation}
For this purpose we find such bound by finding bounds, under fairly general conditions, for the distribution of the maximun of certain partial sums in the following way: let $\xi_1,\xi_2,\cdots,\xi_m$ be random variables; they need not be independent or identically distributed. Let $S_k=\xi_1+\xi_2+\cdots+\xi_k \;\;(S_0=0)$, and put
$$M_m=\max \limits_{0 \leq k \leq m}\left\vert S_k\right\vert.$$
We shall obtain upper bounds for $P\left\{M_m\geq \epsilon\right\}$ by an indirect approach. If

$$M^\prime_m=\max \limits_{0 \leq k \leq m}min\{\left\vert S_k\right\vert,\left\vert S_m-S_k\right\vert\},$$
then
$$M^\prime_m\leq {M_m},$$
it is easy to check that
$$M_m\leq{M^\prime_m+\left\vert S_m\right\vert},$$
and therefore
\begin{equation}
P\left\{M_m\geq \epsilon\right\}\leq P\left\{M^\prime_m\geq {\frac{\epsilon}{2}}\right\} + P\left\{\left\vert S_m\right\vert\geq {\frac{\epsilon}{2}}\right\}.
\label {m}
\end{equation}
If we find separate bounds for the terms on the right in (\ref{m}), we shall have a bound for the term on the left.

We get a bound for the first term via the following lemma:
\begin{lem}
Let us supose that there exists nonnegative numbers $u_1,\cdots,u_m$ such that
$$E\left\{\vert S_j-S_i\vert ^\gamma \vert S_k-S_j \vert ^\gamma \right\}\leq  \left( \sum_{i\leq l\leq j} u_l \right)^\alpha \;\left( \sum_{j\leq l\leq k} u_l \right)^\alpha  ,\;\; 0\leq i \leq j \leq k \leq m   $$
where $\gamma$ are positive and $\alpha\geq \frac{1}{2}$,
then, for all positive $\epsilon$,
$$P\left\{M^\prime_m\geq {\lambda}\right\}\leq \frac {K_{\gamma, \alpha}}{\lambda^{2\gamma}}\;(u_1+u_2+\cdots+u_m)^{2\alpha},$$
where $K_{\gamma, \alpha}$ is a constant depending only on $\gamma$ and  $\alpha$.
\end{lem}

For a proof of the this lemma, see Billingsley (1968, p.89).

Now for a fixed $\delta$ we consider the random variables $\xi_i=U_n(t+\frac{i}{m}\delta)-U_n(t+\frac{i-1}{m}\delta)$, $i=1,\cdots,m$ and for $\gamma=2$ y $\alpha=1$ let us see that:
\begin {equation}
E\{\vert S_j-S_i\vert ^2 \vert S_k-S_j \vert ^2\}\leq   \left(\sum_{i\leq l\leq j} u_l\right) \left(\sum_{j\leq l\leq k} u_l \right) ,\;\; 0\leq i \leq j \leq k \leq m.
\label {los u}
\end {equation}
As $\vert S_j-S_i \vert ^2 = \vert U_n(t+\frac{j}{m}\delta)-U_n(t+\frac{i}{m}\delta)\vert ^2$ and  $\vert S_k-S_j \vert ^2 = \vert U_n(t+\frac{k}{m}\delta)-U_n(t+\frac{j}{m}\delta)\vert ^2$
then
\begin{eqnarray}
\nonumber
&E&\left\{\left\vert U_n(t+\frac{j}{m}\delta)-U_n(t+\frac{i}{m}\delta)\right\vert ^2 \left\vert U_n(t+\frac{k}{m}\delta)-U_n(t+\frac{j}{m}\delta)\right\vert ^2 \right \}\\\nonumber
&=&\frac{1}{n^2}E\left\{\left(\sum\limits_{h=1}^{n}\alpha_h\right)^2 \left(\sum\limits_{h=1}^{n}\beta_h\right)^2\right\}.
\label {5}
\end{eqnarray}

For short we call $\Delta_h=F_h(t+\frac{j}{m}\delta)-F_h(t+\frac{i}{m}\delta)$ and $\gamma_h=F_h(t+\frac{k}{m}\delta)-F_h(t+\frac{j}{m}\delta)$ then $\alpha_h$ takes the value $1-\Delta_h$ with probability $\Delta_h$ if $X_h \in (t+\frac{i}{m}\delta;t+\frac{j}{m}\delta]$, or $-\Delta_h$ with probability $1-\Delta_h$ else.
In the same way $\beta_h$ takes the value $1-\gamma_h$ with probability $\gamma_h$ if $X_h \in (t+\frac{j}{m}\delta;t+\frac{k}{m}\delta]$, or $-\gamma_h$ with probability $1-\gamma_h$ else.
Since the $X_h$ are independent so are the random vector $(\alpha_h, \beta_h)$. Now $E(\alpha_h)=E(\beta_h)=0$ so  (\ref{5}) is equivalent to
$$\frac{1}{n^2}\left\{\sum\limits_{h=1}^{n}E\left(\alpha_h^2\beta_h^2\right) + \sum\limits_{l \ne h}E\left(\alpha_h^2\right)E\left(\beta_l^2\right)+ \sum\limits_{l \geq h}E\left(\alpha_h \beta_h\right)E\left(\alpha_l\beta_l\right)\right\},$$
but

$E(\alpha_h^2)=(1-\Delta_h)^2 \Delta_h + \Delta_h^2(1-\Delta_h) \leq  \Delta_h$,

$E(\beta_h^2) =(1-\gamma_h)^2 \gamma_h + \gamma_h^2(1-\gamma_h) \leq \gamma_h$,

$E(\alpha_h \beta_h)=(1-\Delta_h)(-\gamma_h) \Delta_h - \Delta_h (1-\gamma_h)+ \Delta_h \gamma_h (1-\gamma_h - \Delta_h)=- \Delta_h \gamma_h$,

$E(\alpha_h^2 \beta_h^2)= (1-\Delta_h)^2 \gamma_h^2 \Delta_h + \Delta_h^2 (1-\gamma_h)^2 \gamma_h + \Delta_h^2 \gamma_h^2 (1-\gamma_h - \Delta_h) \leq 2\Delta_h \gamma_h$,
then
\begin{eqnarray*}
\frac{1}{n^2}E\left\{\left(\sum\limits_{h=1}^{n}\alpha_h\right)^2 \left(\sum\limits_{h=1}^{n}\beta_h\right)^2\right\}
&\leq& \frac{1}{n^2}\left \{\sum\limits_{h=1}^{n} 2\Delta_h \gamma_h + \sum \limits_{l\ne h} \Delta_h \gamma_l + \sum \limits_{l\ne h} \Delta_h \gamma_h\Delta_l \gamma_l\right\}\\
&\leq&2\,\sum\limits_{h=1}^{n} \frac{\Delta_h}{n}\sum\limits_{l=1}^{n}  \frac{\gamma_l}{n}\\
&=& \sum\limits_{r=i+1}^{j}u_r\sum\limits_{r=j+1}^{k}u_r,
\end{eqnarray*}
where $u_r=\frac{\sqrt{2\,}}{n}\sum\limits_{h=1}^{n}\left(F_h(t+\frac{r}{m}\delta)-F_h(t+\frac{r-1}{m}\delta)\right).$

Then (\ref {los u}) follows and therefore

\begin{eqnarray*}
P(M_m^\prime \geq \frac{\epsilon}{2})
&\leq&\frac{2^4\,K}{\epsilon^4}\left[\frac{\sqrt{2\,}}{n}\sum\limits_{h=1}^{n}\left(F_h(t+\delta)-F_h(t)\right)\right]^2\\
&\leq&\frac{2^5\,K}{\epsilon^4}\left[\frac{1}{n}\sum\limits_{h=1}^{n}w_h(\delta)\right]^2\\
&\leq&\frac{2^5\,K}{\epsilon^4}C^2(\delta),
\end{eqnarray*}
where $C (\delta)=\limsup_n\frac{1}{n}\sum\limits_{h=1}^{n}w_h(\delta)$ and $K=K_{2,1}$.
From this and (\ref{m}) we have
\begin {equation}
\label{max}
P(M_m \geq \epsilon)\leq \frac{2^5\,K}{\epsilon^4}C^2(\delta)+P\left\{\vert U_n(t+\delta)-U_n(t)\vert \geq \frac{\epsilon}{2}\right\}.
\end{equation}
Now, for each $\omega$, $U_n(s, \omega)$ is right-continuous in s. As $m \to \infty$, therefore, $M_m$ converges to $\sup\limits_{t \leq s \leq t+\delta} \vert U_n(s,\omega)-U_n(t,\omega)\vert$ for each $\omega$. Hence (\ref{max}) implies
\begin{equation}
P\left\{\sup \limits_{t \leq s \leq t+\delta} \vert U_n(s)-U_n(t)\vert \geq \epsilon\right\} \leq \frac{2^5\,K}{\epsilon^4}C^2(\delta)+P\left\{\vert U_n(t+\delta)-U_n(t)\vert \geq \frac{\epsilon}{2}\right\}.
\label{7}
\end{equation}
Because of the asymptotic normality of $\left(U_n(t+\delta),U_n(t)\right)$ for fixed $t$ and $\delta$,

$$U_n(t+\delta)-U_n(t){\stackrel{w}{\longrightarrow}}\, \sigma_U^2\, N, \;\;\;\;\rm {as}\;n \to\infty, $$
with $\sigma_U^2=\left(F(t+\delta)-F(t)\right) \left(1-\left(F(t+\delta)-F(t)\right)\right),$
and then

$$P\left\{\vert U_n(t+\delta)-U_n(t)\vert \geq \frac{\epsilon}{2} \right\} \to P\left\{N\geq \frac{\epsilon}{2\sigma_U}\right\} \;\;\;\;\rm {as}\;n \to\infty.$$

It is easy to check that $\sigma_U^2\leq C(\delta)$, then
$$P\left\{N\geq \frac{\epsilon}{2\sigma_U}\right\}\leq \frac{2^4\sigma_U^4}{\epsilon^2}E\left\{N^4\right\}\leq \frac{2^4 3 C^2(\delta)}{\epsilon^4}.$$

Now, for n exceding some $n_\delta$

$$P\left\{\vert U_n(t+\delta)-U_n(t)\vert \geq \frac{\epsilon}{2} \right\}< \frac{2^4 3 C^2(\delta)}{\epsilon^4}.$$
Hence by (\ref {7})
$$P\left\{\sup \limits_{t \leq s \leq t+\delta} \vert U_n(s)-U_n(t)\vert \geq \epsilon\right\} \leq \frac{2^4\,3}{\epsilon^4}(K+1)C^2(\delta).$$
Given $\epsilon$ and $\eta$, chose $\delta$, since $lim_{\delta \downarrow 0}C(\delta)=0$, so that $\frac{2^4\,3}{\epsilon^4}(K+1)C^2(\delta)\leq \delta \eta$. For $n\geq n_\delta$, follows that
$$P\left\{\sup \limits_{t \leq s \leq t+\delta} \vert U_n(s)-U_n(t)\vert \geq \epsilon\right\} \leq \delta \eta,$$
this complete the proof.
\end{dem}
\label{empirica}

\begin{lem}
Let $\{Z_n\}$ be a sequence of random variables that satisfies
$$Z_n \mathop \rightarrow \limits_{n\to \infty}  0 \;\; {\mbox {a.s.},}$$
then there exists a sequence of numbers $\{a_n\}$ that satisfies  $a_n\mathop \downarrow 0^+$ such that $P(\vert Z_n \vert >a_n) \mathop \rightarrow \limits_{n\to \infty}  0$

\begin {dem}
Let $a_0$ be any real number, and let us define $a_n=a_0$.

As $P(\vert Z_n \vert >\frac{a_0}{2}) \mathop \rightarrow \limits_{n\to \infty}  0$. There exists $n_0$ such that $P(\vert Z_n \vert > \frac{a_0}{2}) < \frac {1}{2}$,  for all $n>n_0$.

So now we define $a_n=\frac{a_0}{2}$ from $n_0$ on and as $P(\vert Z_n \vert >\frac{a_0}{4}) \mathop \rightarrow \limits_{n\to \infty}0$, there exists $n_1$ such that $P(\vert Z_n \vert > \frac{a_0}{4}) < \frac {1}{4}$, for all $n>n_1$.

So now we define $a_n=\frac{a_0}{4}$ from $n_1$ on.

The sequence $a_n$ is defined as

$$ a_n=a_0 \;\;\mbox {if} \;\;n \leq n_0,$$

$$a_n=\frac{a_0}{2}\;\; \mbox {if}\;\; n_0<n \leq n_1,$$

$$a_n=\frac{a_0}{4} \;\;\mbox {if}\;\; n_1<n \leq n_2,$$

and so on.

It is clear that $a_n\mathop \downarrow 0^+$ and $P(\vert Z_n \vert >a_n) \mathop \rightarrow \limits_{n\to \infty}  0$
\end{dem}
\label{lema1}
\end{lem}

\begin {lem}
Let  $f_n, f$ be  real functions such that:\\
\begin{itemize}
\item [{$i$})] $f_n, f$ are monotonous functions for all $n$, \\
\item [{$ii$})] $f_n(x)\mathop \rightarrow \limits_{n} f(x),$ for all $x\in \mathbb{R}$,\\
\item [{$iii$})] $f_n(+\infty)\mathop \rightarrow \limits_{n} f(+\infty),\:\:f_n(-\infty)\mathop \rightarrow \limits_{n} f(-\infty)$, for all $n$,\\
\item [{$iv$})] f is a continuous and bounded function in $\mathbb{R}$, \\
\end {itemize}
then
$$\sup\limits_{x\in \mathbb{R}^+} \left\vert  f_n(x)-f(x)\right\vert\mathop \rightarrow \limits_{n}  0. $$

\begin {dem}
Without loss of generality  we assume $f_n$ and $f$ increasing.
Let us denote:
$I=f(-\infty)=\inf\limits_{x\in \mathbb{R}^+}f(x)$ and $S=f(+\infty)=\sup\limits_{x\in \mathbb{R}^+}f(x)$.
Then, for all $\epsilon >0$, there exists $a<b$ such that $f(a)<I+\frac{\epsilon}{2},\:f(b)>S-\frac{\epsilon}{2}$.

As  $f$ is continuous and bounded in $\lbrack a,b \rbrack$ then $f$ is absolutely continuous in  $\lbrack a,b \rbrack$, so $\exists \:\delta>0$ such that if $ x, \:y\: \in\:\lbrack a,b \rbrack, \:\left\vert x-y \right\vert< \delta$ then $\left\vert f(x)-f(y) \right\vert<\frac{\epsilon}{2}$.

Let us choose $ k \in \mathbb{N}$ such that $\delta >\frac{1}{k}$ and let us define
$$x_0=-\infty,\:\:x_1=a,\cdots,\: x_i=a+\frac{i-1}{k}(b-a)\cdots,\:x_{k+1}=b,\:x_{k+2}=+\infty.$$

Let us first note that:
\begin {equation}
\mbox{if}\: x, \:y\: \in\:\lbrack x_i,\:x_{i+1} \rbrack \:\mbox{for some}\: i, \: \left\vert f(x)-f(y) \right\vert<\frac{\epsilon}{2}.\label{1}
\end {equation}

We will consider the following three cases:

\begin {itemize}
\item If $i=0$ then  $ x, \:y\: \in\:( -\infty,\:a \rbrack$ so, as $f$ is monotonous, we have:

$$I<f(x) \leq f(a)<I+\frac{\epsilon}{2},\:I<f(y)\leq f(a)<I+\frac{\epsilon}{2}.$$
So $\left\vert f(x)-f(y) \right\vert=max \{f(x),\:f(y)\}- min\{f(x),\:f(y)\}< I+ \frac{\epsilon}{2}-I=\frac{\epsilon}{2}.$

\item If $i=k+1$ then $ x, \:y\: \in\:\lbrack\:b,\: +\infty)$ so, as $f$ is monotonous, we have:

$$S>f(x) \geq f(b)>S-\frac{\epsilon}{2},\:S>f(y) \geq f(b)>S-\frac{\epsilon}{2}$$
So $\left\vert f(x)-f(y) \right\vert=max \{f(x),\:f(y)\}- min\{f(x),\:f(y)\}<\frac{\epsilon}{2}.$

\item If $1 \leq i \leq k$ then $x, \:y\: \in\:\lbrack a+\frac{i-1}{k}(b-a),\:a+\frac{i}{k}(b-a) \rbrack$ then $ x, \:y\: \in\:\lbrack a,b \rbrack,\:(x-y)\leq\frac{1}{k} < \delta$, so $\left\vert f(x)-f(y) \right\vert<\frac{\epsilon}{2}.$
\end {itemize}

By an other hand, let us note that:
\begin {equation}
\exists \: n(\epsilon)\in \mathbb{N}, \mbox{such that}, \mbox{for all},\, n \:\geq n(\epsilon), \max \limits_{0 \leq i \leq k+2} \left\vert f_n(x_i)-f(x_i) \right\vert<\frac{\epsilon}{2}.\label{22}
\end {equation}

Precisely, by $ii)$ and $iii)$ $f_n(x_i)\mathop \rightarrow \limits_{n} f(x_i)$, for all $\:n \: \mbox{and} \:0 \leq i \leq k+2 $, and as $k$ is finite we have that $ \max \limits_{0 \leq i \leq k+1} \left\vert f_n(x_i)-f(x_i) \right\vert \mathop \rightarrow \limits_{n} 0.$
Now, let us prove that the following inequality holds from (\ref{1}) and (\ref{22}):
\begin{equation}
\sup  \limits_{x \in \mathbb{R}} \left\vert f_n(x)-f(x) \right\vert <\epsilon, \,\,\mbox{for all} \: n \geq n(\epsilon).\label{3}
\end {equation}
For that, let us consider  any $x\in \mathbb{R}$ and let us note that $\left\vert f_n(x)-f(x) \right\vert <\epsilon,$ for all $\:n \geq n(\epsilon).$
But, given $x\in \mathbb{R}$, there exits one and only one $i$, with $0 \leq i \leq k+1$, such that $\: x \in  \lbrack x_i,\:x_{i+1} \rbrack$ then, by monotony we have that:
$$f_n(x)-f(x) \leq f_n(x_{i+1})-f(x)=f_n(x_{i+1})-f(x_{i+1})+ f(x_{i+1})-f(x),$$
so,  by (\ref{1}) and (\ref{22}) we have:
$$f_n(x)-f(x) \leq \left\vert f_n(x_{i+1})-f(x_{i+1})\right\vert + \left\vert f(x_{i+1})-f(x) \right\vert < \frac{\epsilon}{2}+ \frac{\epsilon}{2}=\epsilon.$$
By an other hand,
$$f(x)-f_n(x) \leq f(x)-f_n(x_i)=f(x)-f(x_i)+ f(x_i)-f_n(x_i),$$
so,  by (\ref{1}) and (\ref{22}) we have:
$$f(x)-f_n(x) \leq \left\vert f(x)-f(x_i)\right\vert + \left\vert f(x_i)-f_n(x_i) \right\vert < \frac{\epsilon}{2}+ \frac{\epsilon}{2}=\epsilon.$$
Then, we have that :
$$f(x)-f_n(x) \leq \epsilon \:\: \mbox{and}\:\: f_n(x)-f(x) \leq \epsilon, \: \mbox{for all}\,\, n \geq n(\epsilon),$$
so $\left\vert f_n(x)-f(x) \right\vert < \epsilon, \: \mbox{for all}\,\, n \geq n(\epsilon)$ and (\ref{3}) follows.
\end{dem}
\label{lema2}
\end{lem}

\begin{cor}
Let $\{Z_n\}$ be a  sequence of random variables with distribution functions $F_n$. We assume $Z_n\mathop \rightarrow \limits_{n\to \infty}^w Z$ and let $F$ be the distribution function of $Z$,  continuous, then
$$\sup_{x\in \mathbb{R}} \vert F_n(x)-F(x) \vert \mathop \rightarrow \limits_{n\to \infty}0.$$
\end {cor}

\section*{Acknowledgements}
The authors express their gratitude to Dr. Gonzalo Perera for highly valuable suggestions.

\thebibliography\\

\bibitem{billingsley} Billignsley, P. (1968). {\it Convergence of Probability Measures}. Wiley, New York.
\bibitem{falcon} Falc\'on,C. \& Perera, G. (2000). {\it Fitting mean transition kernels for evolution models}. (pre-print).
\bibitem{guyon} Guyon, X.(1995){\it Random Field on a Network: Modelling, Statistics, and Applications.} Springer-Verlag.
\bibitem{perera1} Perera, G. (1997). Geometry in $\mathbb{Z}^d$ and the Central Limit Theorem for weakly dependent ramdom fields. {\it J. Theoret. Probab.} Vol.{\bf 10}, No.3, 581-603.
\bibitem{perera2} Perera, G. (1997). Applications of Central Limit Theorem Over Asymptotically Measurable Sets: Regression models. {\it C. R. Acad. Sci. Paris.} t.{\bf 324}, Série I, p.1275-1280.
\bibitem{perera3} Perera, G. (1998). Random fields over lattices and irregular sets (to appear in the Proceedings of the "Workshop on statistical inference for spatial processes", Centre de Recherches Mathématiques, Université de Montreal, Springer-Verlag).
\bibitem{rao} Ranga Rao, R.(1962). Relations Between Weak and Uniform Convergence of Measures with Applications {\it The Annals of Mathematical Statistics} Vol. {\bf 33}, No. 2, 659-680.
\bibitem{serflin} Serfling, R.J. (1980). {\it Approximation Theorems of Mathematical Statistics.} Willy \& Sons.
\end{document}